\documentclass[prl,twocolumn]{revtex4}
\usepackage{graphicx}
\def\gsim{\mathop {\vtop {\ialign {##\crcr 
$\hfil \displaystyle {>}\hfil $\crcr \noalign {\kern1pt \nointerlineskip } 
$\,\sim$ \crcr \noalign {\kern1pt}}}}\limits}
\def\lsim{\mathop {\vtop {\ialign {##\crcr 
$\hfil \displaystyle {<}\hfil $\crcr \noalign {\kern1pt \nointerlineskip } 
$\,\,\sim$ \crcr \noalign {\kern1pt}}}}\limits}
\begin{document}

\title{
Comment on arXiv:0811.1575 entitled \\ ``Quantum phase transitions in the Hubbard model on triangular lattice" \\ by T. Yoshioka, A. Koga and N. Kawakami 
}

\author{Shinji Watanabe$^{1}$, Takahiro Mizusaki$^{2}$,  and Masatoshi Imada$^{1}$}

\affiliation{$^{1,}$Department of Applied Physics, University of Tokyo, Hongo, 
Bunkyo-ku, Tokyo, 113-8656, Japan \\
$^{2}$Institute of Natural Sciences, Senshu University, 
Kanda, Chiyoda, Tokyo 101-8425, Japan}

\begin{abstract}
We show that the phase boundary between the paramagnetic metal and 
the nonmagnetic Mott insulator for the Hubbard model on a triangular lattice
obtained by Yoshioka {\it et al.} in arXiv:0811.1575 does not correctly
represent that of the thermodynamic limit but is an artifact of the 6 by 6
lattice they rely on. After the system size extrapolation, the phase boundary
is located at $U/t \sim 5.2$ as proposed by Morita {\it et al}., J. Phys. Soc. Jpn. {\bf 71} (2008) 2109 and in contrast to Yoshioka {\it et al.}  Here, $U$ is the onsite Coulomb repulsion and $t$ is the nearest-neighbor transfer.
\end{abstract}

\maketitle
Recently, Yoshioka, Koga and Kawakami uploaded a paper entitled "Quantum Phase Transitions in the Hubbard Model on Triangular Lattice" on arXiv:0811.1575.\cite{Yoshioka}
They have discussed phase diagram of the Hubbard model on a triangular lattice
in the ground state as a function of the onsite Coulomb repulsion $U$ and the nearest-neighbor hopping $t$ by using the path-integral renormalization group (PIRG) method.
In particular, they have reproduced the existence of the nonmagnetic insulating phase 
near the Mott transition in agreement with the previous PIRG result by Morita {\it et al.}~\cite{Morita} 
However, they found a phase boundary between a paramagnetic metal and 
a nonmagnetic insulator at $U/t \sim 7.7$ with a large jump of the double occupancy $D=\langle n_{i,\uparrow}n_{i,\downarrow}\rangle$, $\Delta D \sim 0.07$, quantitatively
contradicting the result of Ref.\onlinecite{Morita}, where the phase boundary has been found at $U/t \sim 5.2$ with a small jump $\Delta D$ (less than 0.025) of the double occupancy suggesting a weak first-order or continuous Mott transition. Here $n_{i,\sigma}$ is the number operator at site $i$ with spin $\sigma$. Yoshioka {\it et al.} have claimed that their $``$new" algorithm of iterative method allows going beyond the possible trap in the metastable state. According to them, this makes the difference from the result by Morita {\it et al}. 

Here, we show that the difference does not result from the issue of the trapping but from size effects, and the result by Yoshioka {\it et al.}\cite{Yoshioka} is an artifact of their calculation performed only for $6 \times 6$ lattice. 
We have indeed rechecked the data obtained in the calculation by Morita {\it et al.} and confirmed that the data for the $6 \times 6$ lattice are essentially consistent with the properties of the ground state by Yoshioka {\it et al.}. 
Yoshioka {\it et al.} claimed that their data have better reliability because they have implemented an improved procedure in the iteration of the path integral operation, where more independent paths of the Stratonovich-Hubbard variables are considered.  In fact, although it has not explicitly been described, we have already implemented in the calculations presented in Ref. \onlinecite{Morita} essentially the same algorithm of their iterative algorithm as a derivative one of our standard procedure of PIRG.  This minor algorithmic extension was also supplemented by further implementation of the adaptive interval of the imaginary time slice {\it etc}.  

\begin{figure}[ht]
  \begin{center}
    \includegraphics[width=7cm]{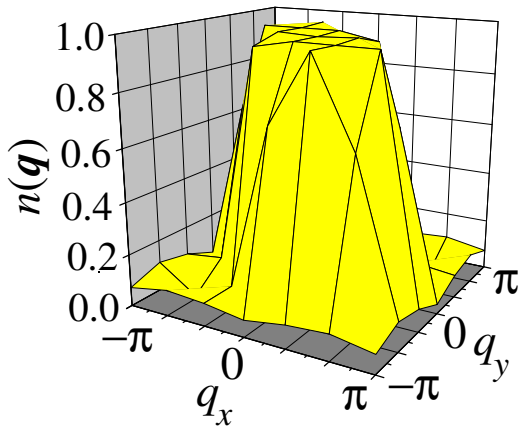}
  \end{center}
  \caption{Momentum distribution $n({\bf q})$ of the ground state at $U/t$=6.5 for 6 $\times$ 6 triangular lattice at half filling.}
  \label{6times6}
\end{figure}

\begin{figure}[ht]
  \begin{center}
    \includegraphics[width=7cm]{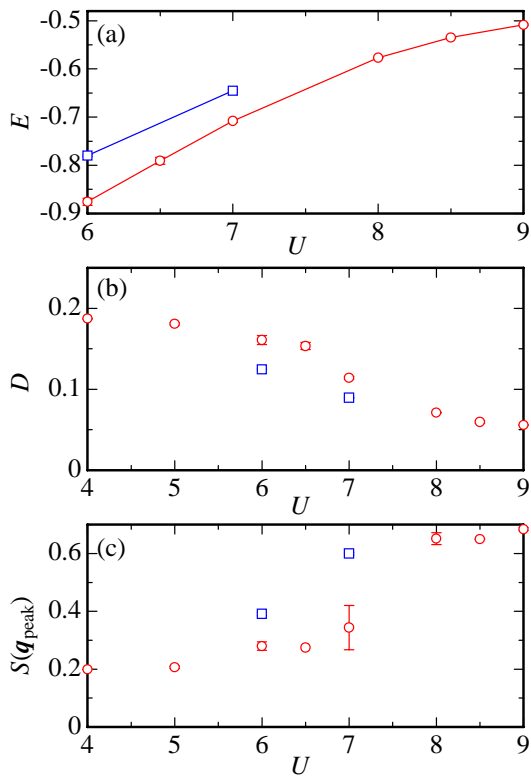}
  \end{center}
  \caption{$U$ dependence of (a) total energy per site $E$ (b) double occupancy $D$ and (c) peak value of $S({\bf q})$ for the ground state (red open circles ) and a typical excited state (blue open squares) of $6 \times 6$ lattice of triangular lattice at half filling.}
  \label{Udependence}
\end{figure}

The ground state energy of the $6 \times 6$ lattice at $U=6.5$ and $t=1$ is $-0.791\pm 0.007$ per site and the double occupancy $D = \langle n_{i\uparrow}n_{i\downarrow}\rangle $ is $0.153 \pm 0.004$ in our calculation. In Fig. \ref{6times6}, we show our data for momentum distribution $n({\bf q})$. These are obtained after the quantum number projection of the total spin.\cite{Mizusaki}
The results show good consistency with the results by Yoshioka {\it et al.} 
In addition, the metastable state shown by Yoshioka {\it et al.} in the range of $6<U<7$ is also consietent with the metastable states we obtained, which have not been discussed in the paper by Morita {\it et al.}, because these are simply excited states. 
The data are also consistent at $U$ larger than 8, where the nonmagnetic insulating state seems to be well stabilized.  In the interval $7<U<8$, our results show that only the state with a nonmagnetic and insulating character with relatively small $D$ around 0.1 is stabilized in contrast to the result by Yoshioka {\it et al.} 
In Fig.\ref{Udependence}, we show the $U$ dependences of the energy $E$, double occupancy $D$ and the spin structure factor $S({\bf q})$ at the peak, both for the ground state and the excited state, showing the behaviors similar to those by Yoshioka {\it et al.} aside from the small discrepancy mentioned above.
Here $S({\bf q})$ is the Fourier transform of the spin correlation 
$\langle (n_{i,\uparrow}-n_{i,\downarrow})(n_{j,\uparrow}-n_{j,\downarrow})\rangle /4$.
The results for the $ 6 \times 6 $ lattice have, therefore, overall consistency between these two independent calculations except for a small discrepancy in the region $7<U<8$.  

A problem of the small size system such as the $6 \times 6$ lattice is that it is not clear what type of phase an obtained state belongs to.  Actually, the ground state we obtained shows the momentum distribution illustrated in Fig.\ref{6times6}, which could be interpreted both by a metal and an insulator in the range of the interaction $6<U<8$.  In any case, for the $6 \times 6$ lattice, $D$ and $S({\bf q})$ seem to show a sharp crossover or level crossing between $U=7$ and 8. This transition was identified as the metal-insulator transition by Yoshioka {\it et al.} 
However, as noted above, the distinction between metals and insulators is not definitely clear when one considers only one size of lattice, and the size scaling is imperative. Actually in our experience, it is hard to distinguish metals from insulators and antiferromagnetic states from paramagnetic ones from the data of a single system size.  For example, if we see Figs. 8, 10, 11 and 13 in Ref.\onlinecite{Kashima} for the square lattice Hubbard model with the next-nearest neighbor transfer, and if we knew only the results up to the $6 \times 6$ lattice, it would not be possible to determine the phase boundaries. 

\begin{figure}[ht]
  \begin{center}
    \includegraphics[width=7cm]{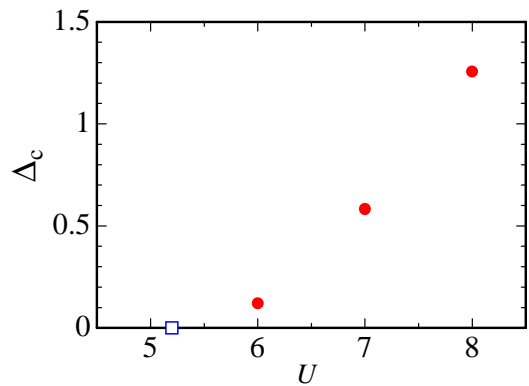}
  \end{center}
  \caption{$U$ dependence of charge gap in the thermodynamic limit 
(red filled circles). Open blue square is located at 
$U=5.2$ where the double occupancy shows a sharp 
change~\cite{Morita} (see Fig.~\ref{double10}).}
  \label{chargegap}
\end{figure}

\begin{figure}[ht] 
\begin{center}
  \includegraphics[width=7cm]{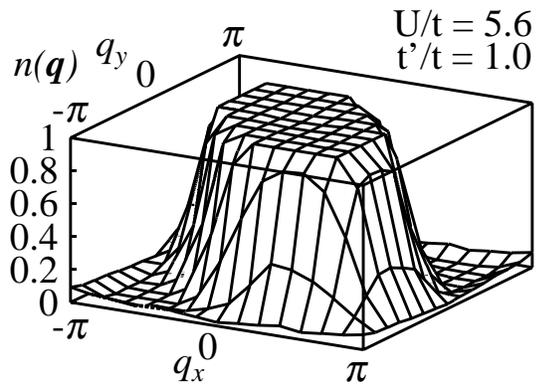}
 \end{center}
\caption{Momentum distribution on $14\times 14$ lattice at
 half-filling for Hubbard model on triangular lattice at $U/t=5.6$}
\label{momdis10}
\end{figure}

In addition, small-sized systems such as $4 \times 4$ and $6\times 6$ often show behaviors very different from larger size systems. This was already cautioned by Kashima {\it et al.}\cite{Kashima}, and Mizusaki {\it et al.}\cite{Mizusaki}, where a clear nonmagnetic insulating phase does not show up for the system size equal to or smaller than $6 \times 6$ for the square lattice Hubbard model with the next-nearest neighbor transfer, while it exists in the system larger than $6 \times 6$ and the size extrapolation supports that it survives in the thermodynamic limit.
This is equally true for the triangular lattice. 
In Fig.\ref{Udependence}, we see a sharp crossover (or level crossing) between $U=7$ and 8 for the $6 \times 6$ lattice.
However, when the system size is increased, the sharp crossover for the $6 \times 6$ lattice diappears already at $8 \times 8$ and the transition is identified at much smaller $U$. In fact, the phase boundary becomes stable among  these larger-sized systems.  The system size dependence of the charge gap $\Delta_c$ defined by $(E_{n+2}+E_{n-2}-2E_{n})/2$ shown in Fig. \ref{chargegap} clearly shows that the region $5.2 <U$ has a nonzero charge gap at half filling.  
Here $E_n$ is the ground state enegy of the $n$ particle systems. Accordingly, the momentum distribution shows insulator-type smooth feature without a jump at the Fermi level\cite{MoritaThesis} as taken from the data in the same calculation with Morita {\it et al.}\cite{Morita} shown in Fig.\ref{momdis10}.
It is unlikely that the obtained states for the larger-sized systems are trapped in the metastable state, because essentially the same state is obtained after the expansion of the Hilbert space even when we start from different initial configurations whichever they are metallic or insulating Hartree-Fock states. During the process of the increasing basis states with iterations, the states converge to similar states.  

\begin{figure}[hptb]
 \begin{minipage}{.49\linewidth}
  \includegraphics[width=\linewidth,clip]{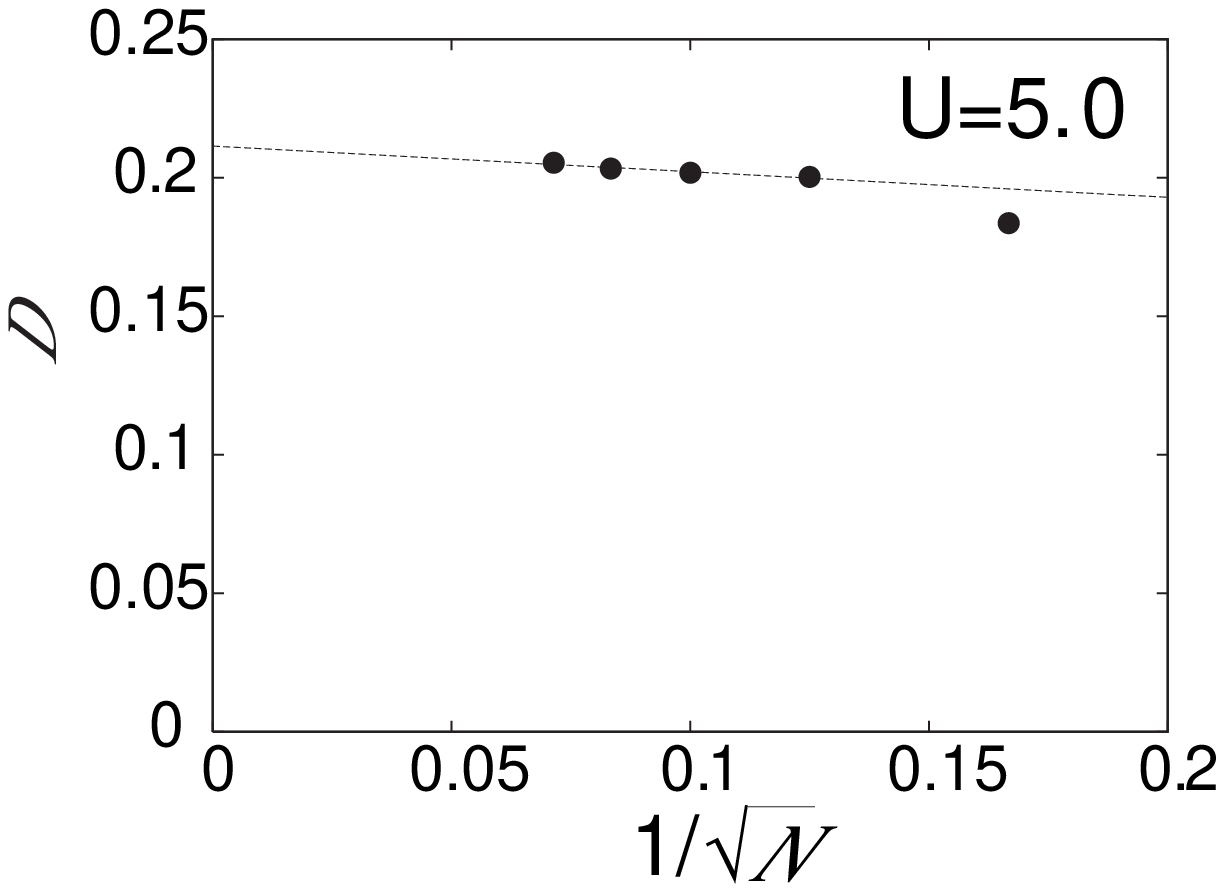}
 \end{minipage}
 \begin{minipage}{.49\linewidth}
  \includegraphics[width=\linewidth,clip]{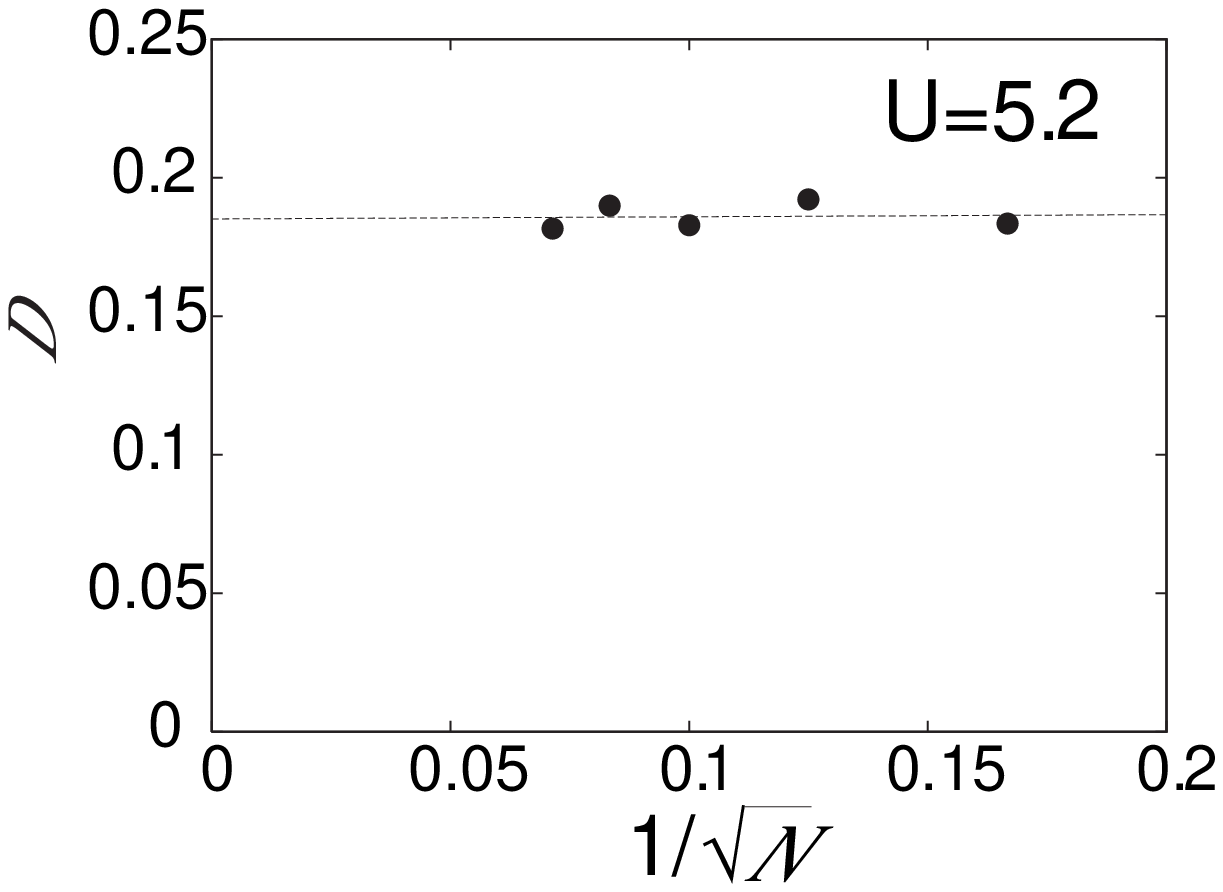}
 \end{minipage}\\
 \begin{minipage}{.49\linewidth}
  \includegraphics[width=\linewidth,clip]{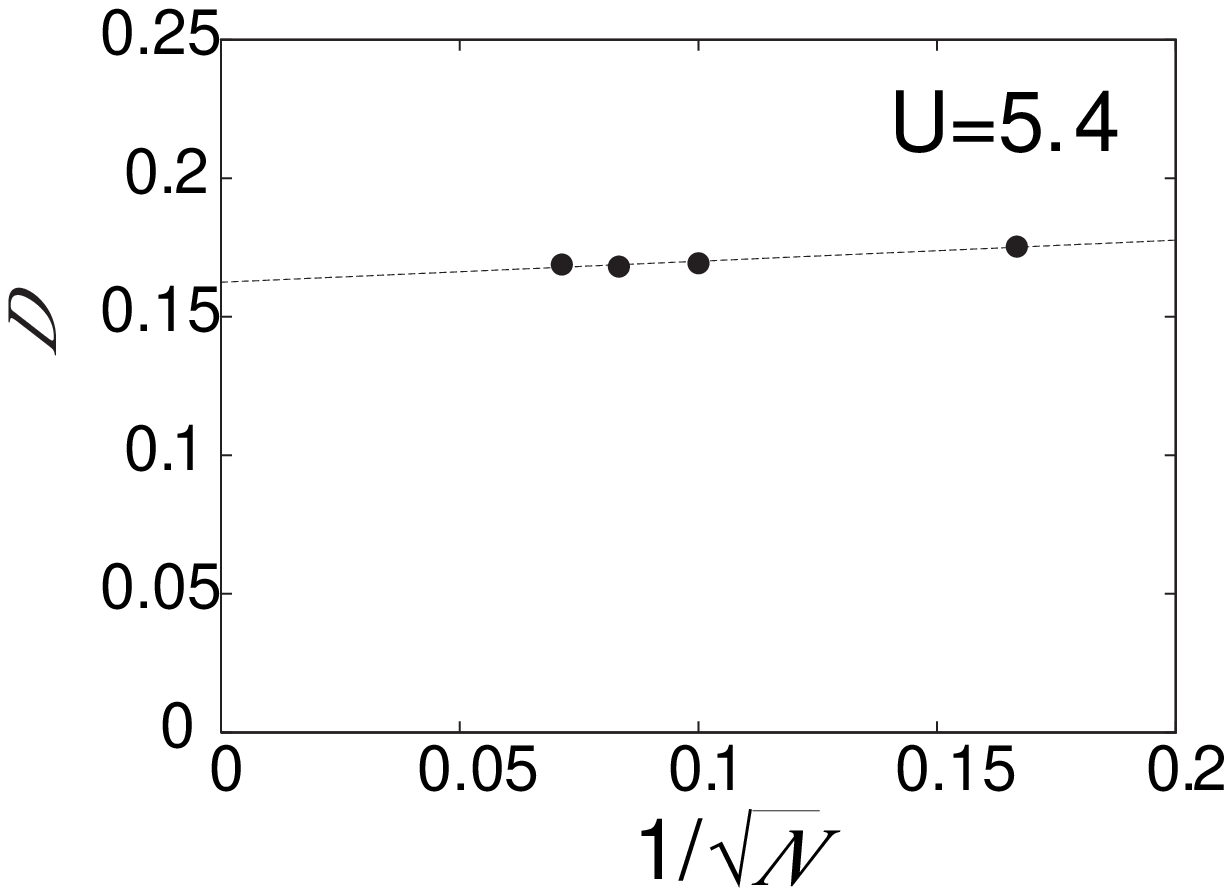}
 \end{minipage}
 \begin{minipage}{.49\linewidth}
  \includegraphics[width=\linewidth,clip]{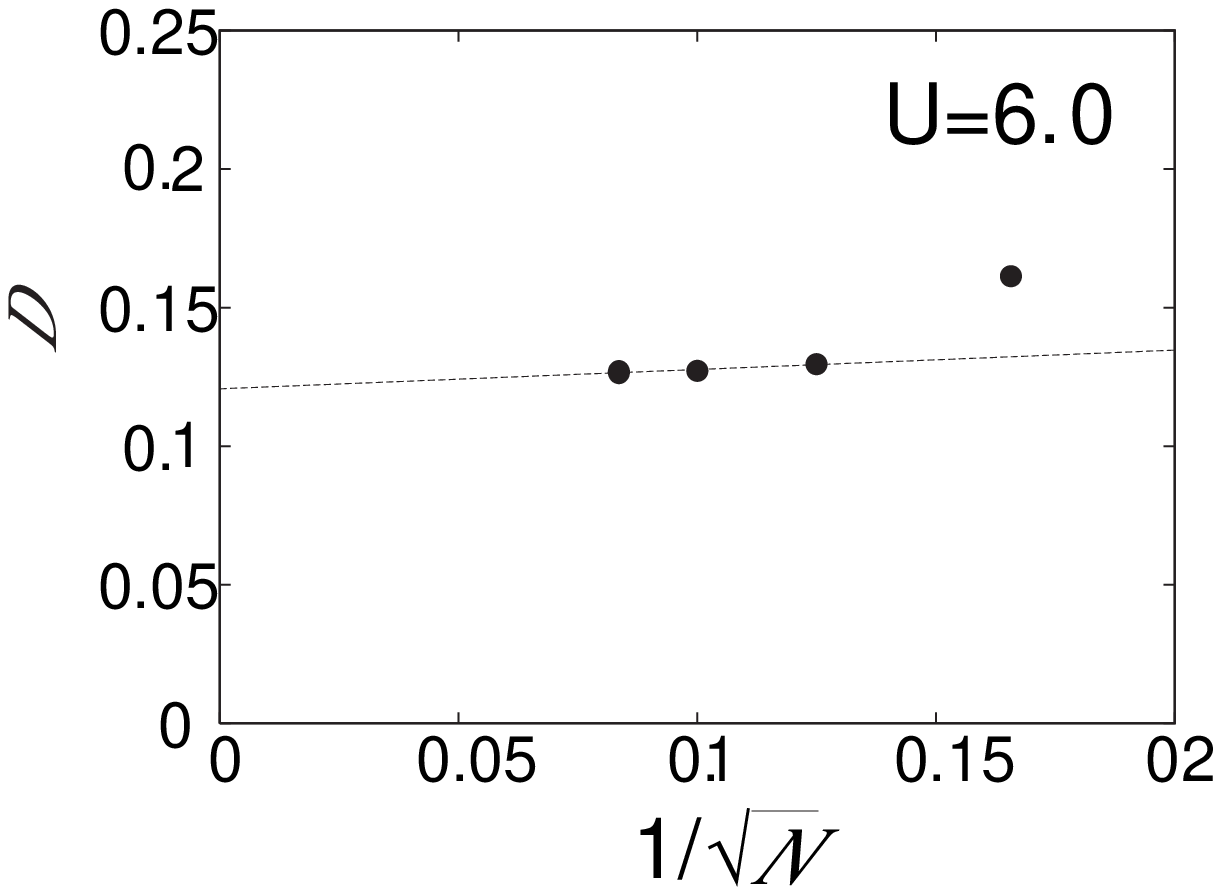}
 \end{minipage}
 \caption{System size dependence of double occupancy $D$ for the triangular lattice at half filling for several choices of $U$. Here the number of sites is $N$.}
 \label{extradob10}
\end{figure}

\begin{figure}
  \begin{center}
    \includegraphics[width=0.8\linewidth,clip]{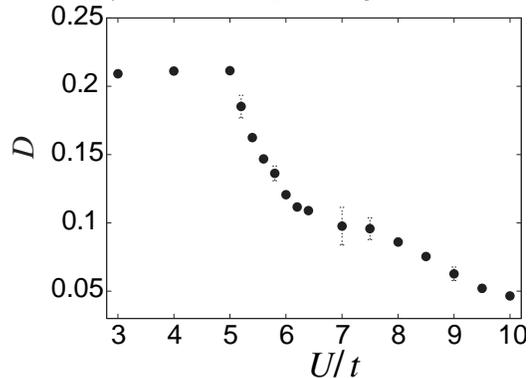}
  \end{center}
  \caption{Double occupancy $D$ in the thermodynamic limit after finite-size
    scaling for triangular lattice at half filling.}  
  \label{double10}
\end{figure}

System size dependences of the double occupancy $D$ for several different choices of $U$ plotted in Fig. \ref{extradob10} reveal\cite{MoritaThesis} that, though $D$ stays above 0.16 up to $U/t=6.0$ for the $6 \times 6$ lattice, the extrapolated values in the thermodynamic limit continuously decrease in the insulating states above $U/t>5.2$ even to $D\sim 0.12$ at $U/t=6$, while $D$ below $U/t = 5.2$ changes its system size dependence to an increasing function with increasing system sizes.  This leads to $D$ higher than 0.2 even just below the Mott transition point $U/t=5.2$ as we see in Fig.\ref{double10}.\cite{Morita,MoritaThesis}   This high value of $D$ in the metallic states near the Mott transition has been already discussed by Tahara {\it et al.} in the case of the square lattice with the next nearest neighbor transfer.\cite{Tahara}  The jump at the Mott transition appears to be small as we see in Fig.\ref{double10}, suggesting a weakly first order or continuous transition.   

From the system size dependence and $U/t$ dependence of the charge gap $\Delta_c$, the momentum distribution $n({\bf q})$, and the spin structure factor $S({\bf q})$, our conclusion for the phase boundary between the paramagnetic metal and the nonmagnetic insulator identified around $U=5.2$ with a weakly first-order character or continuous transition proposed by Morita {\it et al.}\cite{Morita} is solid and we conclude that the results by Yoshioka {\it et al} do not represent the phase boundary of the thermodynamic limit.

At larger $U$, the triangular lattice system may undergo a further transition to some type of antiferromagnetic order. This possibility has not been seriously considered by Morita {\it et al.}\cite{Morita} and calls for further studies. Yoshioka {\it et al.} claimed the transition to 120$^\circ$-type antiferromagnetic order at $U\sim 10$.  However, this possible phase boundary has also to be examined with a careful analysis of finite size effects together with other possibility of antiferromagnetic order with longer-ranged period such as an up-up-down-down structure as proposed in the case of the square lattice Hubbard model with the next-nearest neighbor transfer.\cite{Mizusaki} This is an intriguing open question further to be studied.


\end{document}